\documentclass[8.5pt,twoside,twocolumn]{article}
\usepackage[super,sort&compress,comma]{natbib} 
\usepackage{mhchem}
\usepackage{times,mathptm}
\usepackage{sectsty}
\usepackage{balance} 

\usepackage{graphicx} 
\usepackage{lastpage}
\usepackage[format=plain,justification=raggedright,singlelinecheck=false,font=small,labelfont=bf,labelsep=space]{caption} 
\usepackage{fancyhdr}
\usepackage{color}

\begin{document}

%
%
\thispagestyle{plain}
\fancypagestyle{plain}{
\fancyhead[L]{\includegraphics[height=8pt]{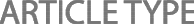}}
\fancyhead[R]{\includegraphics[height=10pt]{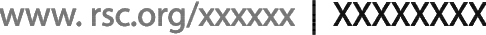}\vspace{-0.2cm}}
\renewcommand{\headrulewidth}{1pt}}
\renewcommand{\thefootnote}{\fnsymbol{footnote}}
\renewcommand\footnoterule{\vspace*{1pt}%
\hrule width 3.4in height 0.4pt \vspace*{5pt}} 
\setcounter{secnumdepth}{5}

\makeatletter 
\def\subsubsection{\@startsection{subsubsection}{3}{10pt}{-1.25ex plus -1ex minus -.1ex}{0ex plus 0ex}{\normalsize\bf}} 
\def\paragraph{\@startsection{paragraph}{4}{10pt}{-1.25ex plus -1ex minus -.1ex}{0ex plus 0ex}{\normalsize\textit}} 
\renewcommand\@biblabel[1]{#1}            
\renewcommand\@makefntext[1]%
{\noindent\makebox[0pt][r]{\@thefnmark\,}#1}
\makeatother 
\renewcommand{\figurename}{\small{Fig.}~}
\sectionfont{\large}
\subsectionfont{\normalsize} 

\fancyfoot{}
\fancyfoot[LO,RE]{\vspace{-7pt}\includegraphics[height=9pt]{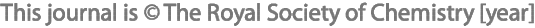}}
\fancyfoot[CO]{\vspace{-7.2pt}\hspace{12.2cm}\includegraphics{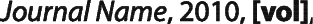}}
\fancyfoot[CE]{\vspace{-7.5pt}\hspace{-13.5cm}\includegraphics{RF.ps}}
\fancyfoot[RO]{\footnotesize{\sffamily{1--\pageref{LastPage} ~\textbar  \hspace{2pt}\thepage}}}
\fancyfoot[LE]{\footnotesize{\sffamily{\thepage~\textbar\hspace{3.45cm} 1--\pageref{LastPage}}}}
\fancyhead{}
\renewcommand{\headrulewidth}{1pt} 
\renewcommand{\footrulewidth}{1pt}
\setlength{\arrayrulewidth}{1pt}
\setlength{\columnsep}{6.5mm}
\setlength\bibsep{1pt}

\twocolumn[
  \begin{@twocolumnfalse}
\noindent\LARGE{\textbf{Microscopic basis for pattern formation and anomalous transport in two-dimensional active gels$^{\dag}$}}
\vspace{0.6cm}

%
%
\noindent\large{\textbf{David A. Head,\textit{$^{a,b,c}$} Gerhard Gompper,\textit{$^{a,d}$} and W. J. Briels\textit{$^{b,e}$}}}\vspace{0.5cm}

%
%
\noindent\textit{\small{\textbf{Received Xth XXXXXXXXXX 20XX, Accepted Xth XXXXXXXXX 20XX\newline
First published on the web Xth XXXXXXXXXX 200X}}}

\noindent \textbf{\small{DOI: 10.1039/b000000x}}
\vspace{0.6cm}

%
%
\noindent \normalsize{Active gels are a class of biologically-relevant material containing embedded agents that spontaneously generate forces acting on a sparse filament network. {\em In vitro} experiments of protein filaments and molecular motors have revealed a range of non-equilibrium pattern formation resulting from motor motion along filament tracks, and there are a number of hydrodynamic models purporting to describe such systems. Here we present results of extensive simulations designed to elucidate the microscopic basis underpinning macroscopic flow in active gels. Our numerical scheme includes thermal fluctuations in filament positions, excluded volume interactions, and filament elasticity in the form of bending and stretching modes. Motors are represented individually as bipolar springs governed by rate-based rules for attachment, detachment and unidirectional motion of motor heads along the filament contour. We systematically vary motor density and speed, and uncover parameter regions corresponding to unusual statics and dynamics which overlap but do not coincide. The anomalous statics arise at high motor densities and take the form of end-bound localized filament bundles for rapid motors, and extended clusters exhibiting enhanced small-wavenumber density fluctuations and power-law cluster-size distributions for slow, processive motors. Anomalous dynamics arise for slow, processive motors over a range of motor densities, and are most evident as superdiffusive mass transport, which we argue is the consequence of a form of effective self-propulsion resulting from the polar coupling between motors and filaments.}
\vspace{0.5cm}
 \end{@twocolumnfalse}
  ]

%
%
\section{Introduction}

%
%
\footnotetext{\dag~Electronic Supplementary Information (ESI) available: [details of any supplementary information available should be included here]. See DOI: 10.1039/b000000x/}

\footnotetext{\textit{$^{a}$~Theoretical Soft Matter and Biophysics, Institut f\"ur Festk\"orperforschung, Forschungszentrum J\"ulich 52425, Germany}}
\footnotetext{\textit{$^{b}$~Computational Biophysics, University of Twente, 7500 AE Enschede, The Netherlands}}

\footnotetext{\textit{$^{c}$~E-mail: d.head@fz-juelich.de}}
\footnotetext{\textit{$^{d}$~E-mail: g.gompper@fz-juelich.de}}
\footnotetext{\textit{$^{e}$~E-mail: w.j.briels@tnw.u-twente.nl}}


%
%
\label{s:intro}

Living matter fundamentally differs from dead (or passive) media in that it is driven by spontaneously-activating internal processes, depleting some energy reservoir maintained by the organism's metabolism\cite{AlbertsBook,Liverpool2006,Ramaswamy2010}. This should be contrasted with passive materials which may be driven externally by {\em e.g.} an imposed boundary, or simply agitated by thermal noise. An immediate and crucial consequence with regards quantitative modeling of living systems is that they do not obey the principles of thermodynamic equilibrium\cite{Lau2003,Mizuno2007}, necessitating the development of novel analytical and theoretical principles if such systems are to be understood on the same level as equilibrium matter. Ideally, such a program should proceed {\em via} incremental improvements to theory in tandem with experimental verification and guidance, but the enormous complexity of real organisms eliminates facile comparison with any suitably transparent theory. It is therefore often prudent to treat {\em in vitro} systems of known composition and reduced complexity.

This approach has been applied with some success to the {\em cellular cytoskeleton}, a dynamic scaffolding of protein filaments and associated proteins that contributes to the mechanical, structural and motility properties of eukaryotic cells\cite{AlbertsBook,HowardBook,Kolomeisky2007,BoalBook,BrayBook}. This can be classified as an {\em active gel}, both because it contains molecular motors that generate ${\mathcal O}(pN)$ forces on the filaments, and because the filaments themselves can translate or `treadmill' due to different growth rates at either end. It is possible with reconstituted {\em in vitro} active gels to inhibit treadmilling and add permanent biotin-avidin crosslinks between filaments, resulting in a static gel with both thermal and athermal sources of noise, the latter deriving from the action of motor proteins on the network\cite{Mizuno2007,Kiehart1986,Mizuno2008,Mizuno2009}. These athermal noise sources have been shown to violate the fluctuation-dissipation relation\cite{Mizuno2007,Mizuno2008}, categorically placing active gels outside the realm of equilibrium thermodynamics, and the athermal noise spectrum can be related to the properties of motor proteins\cite{MacKintosh2008,Levine2009,Head2010}.

Without permanent crosslinks, the filament network can plastically evolve and macroscopic flow may emerge, resulting in non-trivial pattern formation as observed in {\em in vitro} systems consisting of microtubules and various associated motors\cite{Nedelec1997,Surrey2001,Nedelec2001}, and actin-myosin complexes\cite{Backouche2006}. This richer problem has inspired the development of analytical theories incorporating filament flow and active driving\cite{Kruse2004,Kruse2005,Voituriez2005,Voituriez2006,Marenduzzo2007,Basu2008,Cates2008,Liverpool2003,Ziebert2004,Liverpool2004,Ziebert2005,Liverpool2005,Ahmadi2005,Ruhle2008,Bassetti2000,Lee2001,Kim2003,Sankararaman2004}. One approach extends the hydrodynamic equations of liquid crystals in their nematic phase\cite{deGennesProst} to include active terms (with phenomenological coefficients) obeying the required symmetries\cite{Kruse2004,Kruse2005,Voituriez2005,Voituriez2006,Marenduzzo2007,Basu2008,Cates2008}. These active nematodynamic equations admit stable solutions with line defects in the director field, include cylindrical spirals that rotate due to active processes\cite{Kruse2004}; these were likened to the vortices seen in quasi-2D microtubule experiments\cite{Nedelec1997,Surrey2001}, although no quantitative comparison has yet been made. An alternative approach extends the Smoluchoswki equations for rigid rods\cite{DoiEdwards} to include active driving\cite{Liverpool2003,Ziebert2004,Liverpool2004,Ziebert2005,Liverpool2005,Ahmadi2005,Ruhle2008}, and although these active terms were derived from microscopic considerations, some coarse-graining is still required, resulting in what can be regarded as a mesoscale model.

What is lacking despite this plethora of modelling is a clear picture of the microscopic mechanisms driving the self-organization observed in experiments. Ideally the large-scale equations could be found by coarse-graining a suitable microscopic model, but this would inevitably involve approximations whose validity would need to be assessed. Furthermore, the only coarse-graining attempted so far failed to generate all of the terms expected on symmetry grounds, with implications for the formation of vortices\cite{Liverpool2005}. It is into this state of affairs that simulations can play a key role, permitting as they do full control and access of all microscopic details. Non-Brownian simulations mimicking the microtubule experiments generate asters and apparently vortices\cite{Nedelec1997,Surrey2001,Nedelec2001b,Nedelec2002}, but did not include excluded volume interactions between filaments and thus lacked nematic elasticity, making comparison to the nematodynamic theories\cite{Kruse2004,Kruse2005,Voituriez2005,Voituriez2006,Marenduzzo2007,Basu2008,Cates2008} problematic. Further simplified models without filament growth found no defects\cite{Ziebert2008}. Point-like defects were also claimed in two-dimensional simulations of inelastic rods\cite{Aranson2005,Aranson2006,Ziebert2007,Karpeev2007,Jia2008,Ziebert2009}, but these were somewhat coarse-grained and again provided no microscopic picture for the role played by individual motors.

\begin{figure}
\centerline{\includegraphics[width=8.5cm]{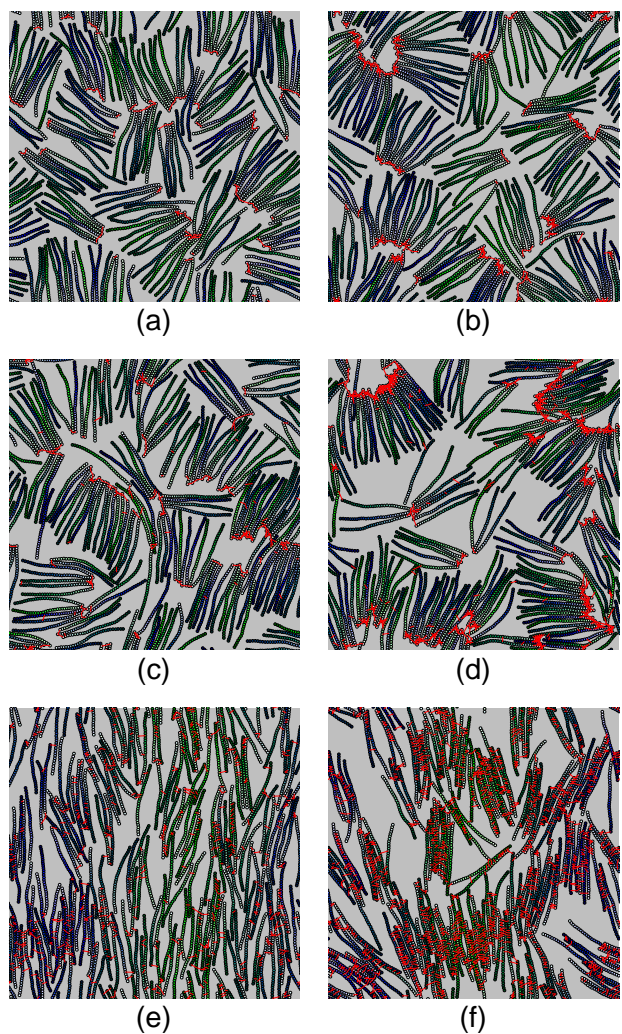}}
\caption{(Color) Snapshots for systems with varying motor density and speeds. The left-hand column (a), (c) and (e) corresponds to a low motor density, and the right-hand column (b), (d) and (f) to high motor density. Similarly, the top row (a), (b) corresponds to fast motors, the middle row (c), (d) to slow, processive motors, and the bottom row (e), (f) to static motors. Filaments are shaded light (dark) towards their $[+]$ ($[-]$) ends, respectively, and the blue-green hue is arbitrary. Motors are shown as red lines. The system size was $4L\times4L$ in all cases. Larger snapshots and movies are available from Ref.\cite{SuppInf}. In terms of the model parameters (see Sec.~\ref{s:model}), $k_{\rm A}/k_{\rm D}=20$ for (a), (c), (e) and 40 for (b), (d), (f); $k_{\rm M}\tau_{b}=75$ for (a), (b), $0.75$ for (c), (d) and $0$ for (e), (f).
}
\label{f:snapshots}
\end{figure}

Here we present the results of extensive simulations of a two-dimensional model for active gels, with the goal of complementing existing analytical and experimental approaches by providing a microscopic underpinning for any observed non-trivial macroscopic pattern formation. Our chosen numerical scheme includes: (i)~Mobile motors as the originator of all non-equilibrium effects; (ii)~Excluded volume interactions between filaments, so nematic elasticity is present; (iii)~Thermal diffusion of filaments, which will be relevant to actin if not microtubule systems; (iv)~Mechanical elasticity of filaments, which can store and release elastic energy in the form of bending and stretching modes. Filaments are polar, that is they have well-defined $[+]$ and $[-]$-ends, and motors move strictly towards $[+]$-ends, breaking microscopic reversibility. Some snapshots are given in Fig.~\ref{f:snapshots}. The structures evident in these snapshots, and associated dynamic anomalies, are fully characterized below. Given the large aspect ratio of single filaments, finite-size effects are pronounced and we have taken great efforts to control variation with system size for all of the quantities considered.

One of our central findings is that the action of the motors can drive the formation of non-trivial structure formation in one region of parameter space, and anomalous dynamics in a different, albeit overlapping parameter regime. Motor-driven structure formation is most evident for high motor densities, where the nematic ordering breaks down and filament clusters form. The nature of these clusters depends on the motor speed. For motors sufficiently fast to dominate over thermal diffusion, localized clusters form in which the filaments are bound at their $[+]$-ends, as evident in Fig.~\ref{f:snapshots}(b). This can be viewed as non-equilibrium polar ordering on scales comparable to the filament length~$L$, that becomes isotropic on much larger lengths. The dynamics of such states is anomalous in that the mean-squared displacement of filament centers exhibits a {\em superdiffusive} regime in which it scales faster than linearly in time, but this effect becomes less pronounced as the motor speed increases, a non-intuituve finding that we explain below in terms of a vanishing population of mobile motors.

Maintaining a high motor density but decreasing the motor speed so that motor motion and thermal effects compete, we find a region of parameter space that exhibits both anomalous diffusion and non-trivial structure formation; see Fig.~\ref{f:snapshots}(d). Extended, polarised clusters form that can span the system size, generating large voids between the high-density clusters that is evident as an increase, possibly a divergence, of the static structure factor $S(q)$ as the wavenumber $q\rightarrow0$. Superdiffusive mass transport is most pronounced in this parameter regime. Nonetheless the polarity of filaments separated by distances comparable to a filament length or more are uncorrelated, in contrast to the prevalent assumption of continuum modelling that the polarity field is slowly-varying over such lengths.

We also identify a third region of parameter space in which the system is structurally similar to the equilibrium limit with no motors, exhibiting nematic ordering similar to Fig.~\ref{f:snapshots}(e), but nonetheless displays anomalous, superdiffusive transport. This regime corresponds to the same motor speeds that generate the extended clusters described above, but a lower density of motors. The key observation underlying this quandary is that although the {\em nematic} ordering of the filaments appears normal, the {\em polarity} of nearby filaments are nonetheless correlated over distances corresponding to a number of filament diameters. Since the coupling between motors and filaments is polar, such correlations can result in a persistent direction of motor forces acting between filaments, generating net translational motion that can be viewed as a form of emergent self-propulsion. This intuitive idea is supported by showing filament velocity and polarity are most strongly correlated for parameters for which anomalous transport reaches its maximum.

This paper is arranged as follows. In Sec.~\ref{s:model} we describe in detail the numerical model, before turning into Sec.~\ref{s:passive} to briefly describe the behavior of passive systems in which motors do not move. This provides a comparison for the results of active systems presented in the succeeding section, where we separately focus on static quantities in Sec.~\ref{s:statics} and then the dynamics in Sec.~\ref{s:dynamics}. Despite performing a systematic investigation of parameter space, and including the physical mechanisms thought to be necessary for the emergence of the structures seen in the microtubule experiments\cite{Nedelec1997,Surrey2001}, we never observe vortices. Possible causes are discussed in Sec.~\ref{s:discussion}.

%
%

%
%
\section{Model}
\label{s:model}

\begin{figure}
\centerline{\includegraphics[width=8cm]{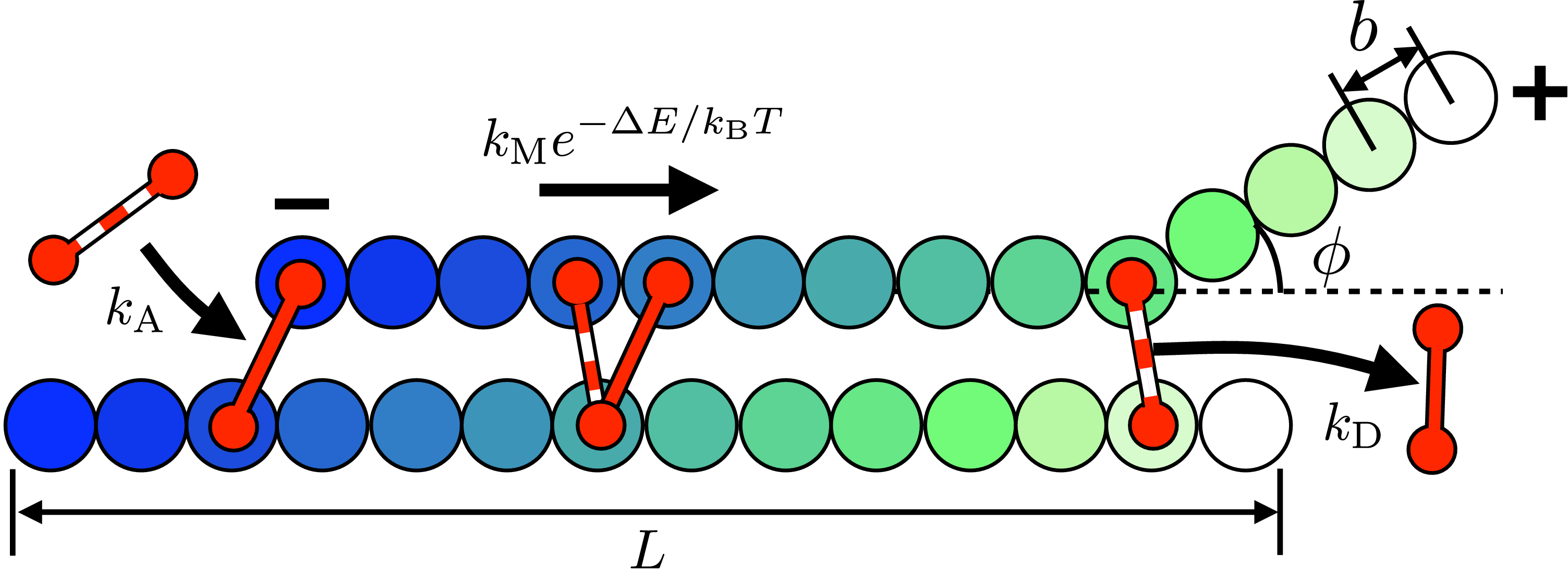}}
\caption{(Color online) Schematic of model parameters. Filament polarity is defined by $[+]$ and $[-]$-ends as denoted in the figure. Motors are 2-headed springs defined by attach, detach and motion rates $k_{\rm A}$, $k_{\rm D}$ and $k_{\rm M}$, respectively, where the movement rate is attenuated by an exponential factor depending on the increase in spring energy $\Delta E$ for the proposed move ($\Delta E=0$ if this change is negative). See text for details.}
\label{f:model_schematic}
\end{figure}

We are interested in determining the universal features of motor-driven filament systems,  independent of the atomic details of either the filaments or the motors (or motor clusters). To this end, we describe both filament segments and motors in a simplified manner requiring only a few degrees of freedom, and focus on the collective effects of many motors and filaments. See Fig.~\ref{f:model_schematic} for a schematic representation of  essential aspects of the model, which we now describe in detail.

Filaments are described as polar semi-flexible polymers of monomers with lattice spacing~$b$. Rigidity and contour length are maintained by elastic energies penalizing local fluctuations in both monomer separation and bending. For the former, two monomers instantaneously separated by a distance $r$ incurs a cost $\delta E^{\rm stretch}=\pi a^{2} E(r-b)^{2}/2b$, where $E$ is the material's Young's modulus and $a$ the radius of the cross-section (assumed circular)\cite{LandauLifschitz}. In addition, the energy for bending by a local angle $\phi$ is given by $\delta E^{\rm bend}=\kappa\phi^{2}/2b$ with $\kappa$ the bending modulus, which for a rod with a circular cross-section of radius $a$ is given by $\kappa=\pi Ea^{4}/4$\cite{LandauLifschitz}. Once thermal fluctuations from the solvent are included (see below), and assuming sufficient resistance to stretching to inhibit contour length fluctuations, this gives a persistence length $\ell_{p}\approx\kappa/2k_{\rm B}T$ in 2 dimensions. For our parameters, $\ell_{p}/L\approx10/3$ in terms of the filament length~$L$, so our filaments are semiflexible.

Excluded volume is incorporated as a repulsive potential between non-bonded monomers. We use the repulsive part of the Lennard-Jones potential with range $\sigma$ and energy~$\varepsilon$, truncated and shifted to ensure continuity of the first derivative at the maximum range $2^{1/6}\sigma$, {\em i.e.} $V^{\rm ev}(r)=4\varepsilon(\sigma/r)^{6}\left[(\sigma/r)^{6}-1\right]+\varepsilon$, with $r$ the distance between monomer centers. To avoid a proliferation of parameters we simply set $\sigma=b$, while $\varepsilon=5k_{\rm B}T$ is set sufficiently high to ensure excluded volume dominates over thermal fluctuations for overlapping monomers.

Thermal fluctuations arise due to the interaction between the filaments and the solvent. Assuming the usual low-Reynolds number limit for biophysical systems, we adopt a Brownian dynamics algorithm that describes forces due to solvent fluctuations in the non-inertial limit\cite{AllenTildesly}. To better model the anisotropic friction coefficients for elongated particles\cite{DoiEdwards}, we first project forces parallel and perpendicular to the local filament axis, and then apply the usual displacement increments with solvent friction coefficients $\gamma^{\parallel}$ and $\gamma^{\perp}=2\gamma^{\parallel}$, respectively. Note that solvent forces and drag obey the usual fluctuation-dissipation relation {\em via} the temperature $k_{\rm B}T$, so the solvent is in equilibrium; all non-equilibrium effects derive from the motion of motors, which we now describe.

Motors (or motor clusters) are modeled as two-headed harmonic springs that are simultaneously attached to two monomers; motors with one or no attached heads are not explicitly represented. The spring extension is defined in terms of the separation between the monomer centers relative to its natural length, here taken to be the excluded volume range~$2^{1/6}\sigma$, and the spring stiffness is fixed at~$k_{\rm B}T/b^{2}$. Attachment, detachment and motor motion are defined by the following rate-based rules. (i)~Motors attach at a rate $k_{\rm A}$ between any two monomers on different filaments whose centers are within a specified distance, which for simplicity we take to be the spring's natural length~$2^{1/6}\sigma$. (ii)~Both motor heads simultaneously detach at a rate $k_{\rm D}$ that does not depend on the positions of heads along the filaments, nor on the spring tension, except in that severely stretched motors with elastic energies exceeding $\approx7.5k_{\rm B}T$ detach immediately. Increased detachment rates at filament ends, thought to be important for the formation of vortices (but not asters)\cite{Surrey2001}, could easily be included at a later stage. (iii)~Each motor head moves in a Monte Carlo-like manner: The change in motor spring energy $\Delta E$ for a head to move $m\geq1$ monomers towards the attached filament's $[+]$--end is calculated, and this trial move is accepted with probability $k_{\rm M}e^{-\Delta E/k_{\rm B}T}$ per unit time if $\Delta E>0$, or $k_{\rm M}$ if $\Delta E<0$. Each $m$ is drawn uniformly from the fixed range $[1,m_{\rm max}]$, and the acceptance probability is suitably normalized to ensure invariance with respect to $m_{\rm max}$.
Motors already in tension are less likely to move due to the increase in strain energy, giving an effective stall force of the order of $k_{\rm B}T/b$. Motors already at $[+]$-ends simply do not move, but may detach at the usual rate. Motors do not move if to do so would exceed the maximum spring energy described in~(ii).

All filaments have $M=30$ monomers and hence are of length $L=30b$, giving an aspect ratio $L/b=30$. Typically we use a $4L\times 4L$ box, but to check for convergence with system size for any given quantity, we also simulated $6L\times6L$ and $8L\times8L$ boxes for a representative sample of parameter space, corresponding to 2 vertical and 2 horizontal lines in the parameter space discussed below. The area fraction of filaments was fixed at $\phi\approx21\%$ throughout, which was checked to give a nematic order parameter close to unity in the absence of motors. At $t=0$ filaments are placed in a smectic configuration to ensure no significant initial overlap, with each filament's polarity independently chosen to be $\pm\hat{\bf p}$ with equal probability. Steady state is identified as when the two--time mean squared displacement of filament centres--of--mass~${\bf x}(t)$, $\langle\Delta r^{2}(t_{\rm w},t_{\rm w}+t)\rangle=\langle|{\bf x}(t_{\rm w}+t)-{\bf x}(t_{\rm w})|^{2}\rangle$, ceases to depend on $t_{\rm w}$ and only varies with the lag time~$t$, {\em i.e.} becomes $\langle\Delta r^{2}(t)\rangle$. 

The results from the simulations will be described in terms of the two dimensionless parameters $k_{\rm A}/k_{\rm D}$ and $k_{\rm M}\tau_{b}$, where $\tau_{b}=Lb\gamma/4k_{\rm B}T$ is the time for an isolated filament's centre--of--mass to translationally diffuse over one monomer diameter. Thus $k_{\rm A}/k_{\rm D}$ gives a rough measure of motor density, and $k_{\rm M}\tau_{b}$ measures the competing effects of motor motion to thermal diffusion over lengths of the monomer spacing~$b$. We fix the value of $k_{\rm D}$ to be small relative to~$\tau_{b}^{-1}$, {\em i.e.} $k_{\rm D}\tau_{b}=7.5\times10^{-3}\ll1$, and consider $k_{\rm M}$ varying from $k_{\rm M}\ll k_{\rm D}$ to $k_{\rm M}\gg k_{\rm D}$, where the former case corresponds to non-processive motors that detach before significant motion, and the latter to processive motors which may move many monomers before detaching.

%
%

%
%
\section{Passive systems: $k_{\rm M}\equiv0$}
\label{s:passive}

We first summarize the behavior of passive systems with $k_{\rm M}=0$, before turning to active systems with $k_{\rm M}>0$. Without motor motion, filament polarity is irrelevant and the time-averaged effect of motor attachment and detachment is to generate an effective attraction between filaments. As the ratio $k_{\rm A}/k_{\rm D}$ between attachment and detachment rates increases, the magnitude of motor-mediated attraction increases and induce a heterogeneous density distribution as evident in Fig.~\ref{f:snapshots}(f). Increasing $k_{\rm A}/k_{\rm D}$ even further gives a percolating network that nonetheless does not fully phase separate over our simulation time.

\begin{figure}
\centerline{\includegraphics[width=8cm]{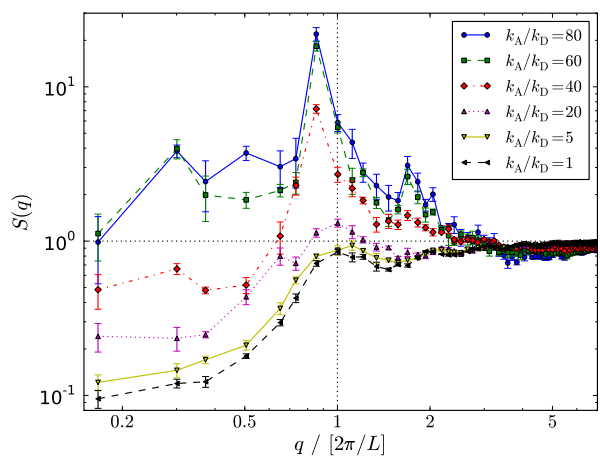}}
\caption{(Color online) Static structure factor $S(q)$ for $k_{\rm M}=0$ and the $k_{\rm A}/k_{\rm D}$ given in the key on log-log axes. The horizontal and vertical dotted lines correspond to $S(q)=1$ and $q=2\pi/L$, respectively, with $L$ the filament length. Error bars give the spread between independent runs. In all cases the system size was $6L\times6L$.}
\label{f:Sq_kM0}
\end{figure}

To identify the length scales associated with the density fluctuations, we calculate the static structure factor $S(q)$ of filament centers, defined as the angular average of \mbox{$S({\bf q})={N}^{-1}\langle|n({\bf q})|^{2}\rangle$} over \mbox{$\hat{\bf q}={\bf q}/q$} with $q=|{\bf q}|$. Here, \mbox{$n({\bf q})=\sum_{\alpha=1}^{N}e^{-i{\bf q}\cdot{\bf x^{\alpha}}}$} is the Fourier-transformed distribution of the filament center-of-masses ${\bf x}^{\alpha}$, $\alpha=1\ldots N$. Results for $1\leq k_{\rm A}/k_{\rm D}\leq 80$ are shown in Fig.~\ref{f:Sq_kM0}. For low $k_{\rm A}/k_{\rm D}$, $S(q)$ exhibits a small peak around the wavenumber corresponding to the filament length~$L$, suggesting a slight degree of smectic ordering with filaments aligned end-to-end, before decaying quadratically for smaller~$q$. Increasing $k_{\rm A}/k_{\rm D}$ to around~$40$, corresponding to ${\mathcal O}(10)$ motors per filament, dramatically enhances the height of this peak. Snapshots such as Fig.~\ref{f:snapshots}(f) reveal local filament bundles that align end-to-end, explaining this peak. For $k_{\rm A}/k_{\rm D}$ in the range 60---80, the $S(q)$ collapse onto a single curve with $S(q)\approx1-3$ as $q\rightarrow0$ and snapshots reveal similar pictures of percolating networks. It is clear that the system does not phase separate for $k_{\rm A}/k_{\rm D}\geq60$ on our simulation times, but instead undergoes kinetic arrest into a long-lived metastable gel. We therefore restrict attention to $k_{\rm A}/k_{\rm D}\leq 40$ for the active systems in Sec.~\ref{s:active}.

\begin{figure}
\centerline{\includegraphics[width=8cm]{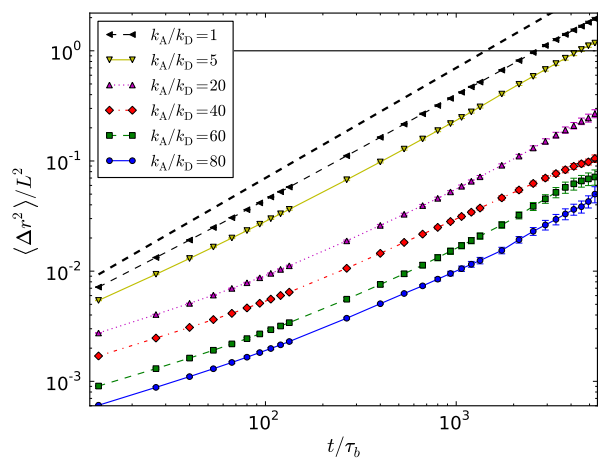}}
\caption{(Color online) Mean squared displacement $\langle\Delta r^{2}\rangle$ for $k_{\rm M}=0$ and $k_{\rm A}/k_{\rm D}$ given in the key for the same systems as Fig.~\ref{f:Sq_kM0}. The thick dashed line has slope~1 corresponding to normal diffusive scaling $\langle\Delta r^{2}\rangle\propto t$, and the solid horizontal line corresponds to displacements equal to the filament length.}
\label{f:msd_kM0}
\end{figure}

One of our key findings for active systems is the existence of enhanced diffusion, so for comparison we present in Fig.~\ref{f:msd_kM0} the mean-squared displacements for passive systems. Given the lack of motor-mediated driving, the sole effect of motors is to bind filaments and thus reduce self-diffusion, and this trend is immediately apparent from the figure. For low motor densities the mass transport is roughly diffusive, with a diffusion constant that decreases with increasing $k_{\rm A}/k_{\rm D}$ and the binding between filaments increases. At high attach rates the mass transport is substantially reduced, and never becomes fully diffusive over our data window, instead giving weak subdiffusion $\langle\Delta r^{2}\rangle\sim t^{\alpha}$ with $\alpha\approx0.9$ at the maximum time lags available. We nonetheless expect normal diffusion with $\alpha=1$ to be recovered at late times.

%
%

%
%

%
%
\section{Active systems: $k_{\rm  M}>0$}
\label{s:active}

The movement of motor heads along the polar filaments generates equal-and-opposite forces that drive relative motion between filaments or filament clusters. Non-trivial flows and pattern formation may thus become stable. In this section we describe the structure and dynamics of active systems in terms of the 2 dimensionless parameters $k_{\rm A}/k_{\rm D}$, which broadly corresponds to the density of motors, and the bare motor speed~$k_{\rm M}\tau_{b}$.

%
%
\subsection{Statics}
\label{s:statics}

Snapshots for systems with mobile motors reveal similar filament ordering as for static motors when $k_{\rm A}/k_{\rm D}$ remains below some crossover value. This crossover value depends on parameters such as filament length and motor spring stiffness, and for the systems considered here occurs around~$k_{\rm A}/k_{\rm D}\approx15$. When $k_{\rm A}/k_{\rm D}$ exceeds this value, the increased motor density induces filaments clustering and non-trivial structure formation, the nature of which depends on the speed of the motors. For $k_{\rm M}\ll k_{\rm D}$, motors detach at a faster rate than moving and become non-processive. Filaments form apolar bundles similar to the passive case~$k_{\rm M}\equiv0$. Conversely, for $k_{\rm M}\tau_{b}\gg1$ motor motion dominates over thermal diffusion and we observe filament clusters bound at their $[+]$-ends as in Fig.~\ref{f:snapshots}(b). For the intermediate regime $k_{\rm D}\tau_{b}\ll k_{\rm M}\tau_{b}\ll 1$, motors are processive but compete with diffusion in structure formation. This results in the extended clusters evident in Fig.~\ref{f:snapshots}(d). 

\begin{figure}
\centerline{\includegraphics[width=8cm]{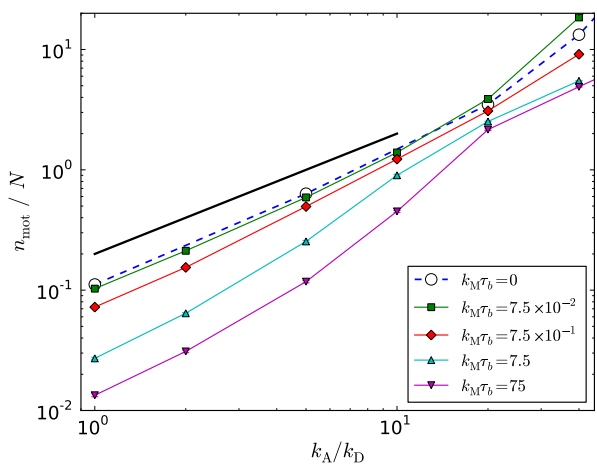}}
\caption{(Color online) Number of motors per filament $n_{\rm M}/N$ versus $k_{\rm A}/k_{\rm D}$ for the dimensionless motor rates $k_{\rm M}\tau_{b}$ given in the key. The thick black line segment has slope~1. For all lines, symbols are larger than the error bars.}
\label{f:motorDensity}
\end{figure}

Underlying the observed clustering is a monotonic increase of the density of attached motors as the ratio $k_{\rm A}/k_{\rm D}$ is increased. As plotted in Fig.~\ref{f:motorDensity}, the number of motors per filament is roughly proportional to $k_{\rm A}/k_{\rm D}$, with a prefactor that decreases with motor speed~$k_{\rm M}$. This linear dependency on $k_{\rm A}/k_{\rm D}$ can be understood as the steady state solution of the simple rate equation \mbox{$\partial_{t}n_{\rm mot}\propto k_{\rm A}-n_{\rm mot}k_{\rm D}$} governing the number $n_{\rm mot}$ of attached motors between two monomers held within the proscribed attachment range. Deviations from strict linearity are evident at high $k_{\rm A}/k_{\rm D}>\mathcal{O}(10)$ when motors induce clustering and the global attachment rate becomes coupled to structure formation. The decrease in motor density for high $k_{\rm M}$ is likely due to the greater fraction of motors stretched close to the detachment limit mentioned in Sec.~\ref{s:model}, but this effect is reduced for higher $k_{\rm A}/k_{\rm D}$ when the densities become similar for all motor speeds.

\subsubsection{Cluster size distributions: }

\begin{figure}
\centerline{\includegraphics[width=8cm]{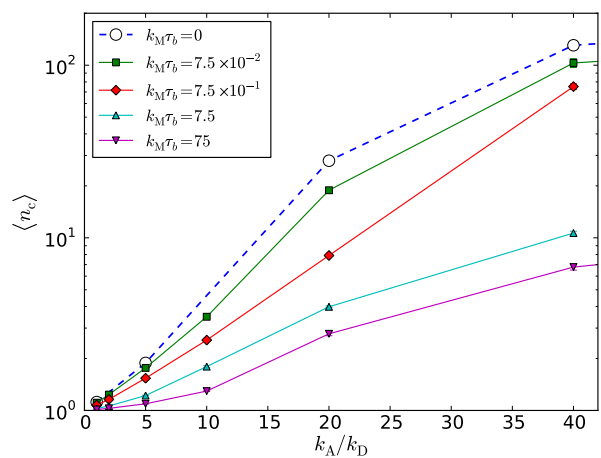}}
\caption{(Color online) Mean number of filaments in a cluster $\langle n_{\rm c}\rangle$ versus $k_{\rm A}/k_{\rm D}$ for the motor speeds given in the key. The system size was fixed at $4L\times 4L$ and error bars are smaller than the symbols.}
\label{f:meanClust}
\end{figure}

To quantify the clustering apparent in snapshots, we first consider the distribution $P(n_{\rm c})$ of clusters consisting of $n_{\rm c}$ filaments, averaged over time and independent runs, where two filaments are considered to belong to the same cluster if there is at least one motor simultaneously attached to both. The mean cluster size $\langle n_{\rm c}\rangle=\sum_{n_{\rm c}} n_{\rm c}P(n_{\rm c})$ is plotted in Fig.~\ref{f:meanClust}, and reveals a monotonic increase with $k_{\rm A}/k_{\rm D}$ following a roughly exponential relationship. It also {\em decreases} with increasing motor speed~$k_{\rm M}$, in part because the density of motors decreases (see Fig.~\ref{f:motorDensity}), and also because they drive the formation of {\em localised} filament clusters bound at their ends. Increasing the system size increases cluster sizes for high $k_{\rm A}/k_{\rm D}$, but the effect is small on logarithmic scales.

\begin{figure}
\centerline{\includegraphics[width=8cm]{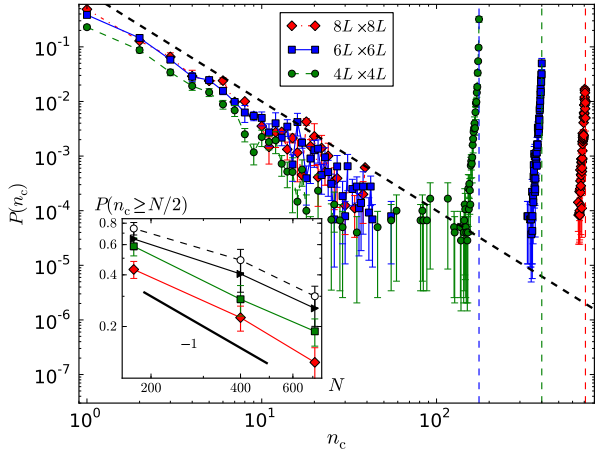}}
\caption{(Color online) Log-log plot of the distribution $P(n_{\rm c})$ of cluster sizes $n_{\rm c}$ for $k_{\rm A}/k_{\rm D}=40$, $k_{\rm M}\tau_{b}=0.075$ and the system sizes given in the key. The thick, dashed diagonal line has a slope of~$-2$, and the vertical lines correspond to $n_{\rm c}=N$ with $N=175$, 400 and 711 the total number of filaments in the system. {\em (Inset)}~$P(n_{\rm c}\geq N/2)$ versus $N$ on log-log axes for (from top to bottom) $k_{\rm M}\tau_{b}=0$, $7.5\times10^{-3}$, $7.5\times10^{-2}$ and $7.5\times10^{-1}$. $P(n_{\rm c}\geq N/2)\equiv0$ for higher $k_{\rm M}\tau_{b}$. The thick black line segment has a slope of~$-1$.}
\label{f:clustDist_kA4em2_kM1em2}
\end{figure}

Considering only the mean cluster size $\langle n_{\rm c}\rangle$ obscures the fact that the shape of the full distribution $P(n_{\rm c})$ qualitatively changes as the parameter space is traversed, see Fig.~\ref{f:clustDist_kA4em2_kM1em2}. For most of the parameter space considered, $P(n_{\rm c})$ is well approximated by a simple exponential decay, and does not vary with system size. However, for high $k_{\rm A}/k_{\rm D}$ and $k_{\rm M}\tau_{b}\stackrel{<}{\scriptstyle\sim}1$, the distribution $P(n_{\rm c})$ deviates from an exponential, in two ways. Firstly, a second peak corresponding to system-spanning clusters with \mbox{$n_{\rm c}\approx N$} emerges for $k_{\rm M}\tau_{b}\stackrel{<}{\scriptstyle\sim}1$. Secondly, and only for slow, processive motors with $k_{\rm D}\tau_{b}\stackrel{<}{\scriptstyle\sim}k_{\rm M}\tau_{b}\stackrel{<}{\scriptstyle\sim}1$, the small-$n_{\rm c}$ decay becomes power law rather than exponential. Although noisy, this algebraic decay can be approximately fitted by an exponent~-2, as shown in Fig.~\ref{f:clustDist_kA4em2_kM1em2}. We note that the region of parameter space for which $P(n_{\rm c})$ exhibits power-law decay coincides with those parameters that exhibit anomalous small-wavenumber density fluctuations, as described in Sec.~\ref{s:Sq}.

The magnitude of the second peak decays monotonically with increasing system size, suggesting it will vanish in the limit of large system size. This is apparent in plots of the integrated area of the second peak, \mbox{$P(n_{\rm c}\geq N/2)=\sum_{n_{\rm c}=N/2}^{N}P(n_{\rm c})$}, which can be understood as the probability to find a cluster that is comparable in extent to the system size. The lower cut-off $N/2$ is arbitrary, but as long as it falls into the middle region where $P(n_{\rm c})\approx0$, as here, its precise choice does not measurably alter the result. As shown in the inset to Fig.~\ref{f:clustDist_kA4em2_kM1em2}, this quantity decays roughly as~$N^{-1}$. $P(n_{\rm c}\geq N/2)$ also monotonically decreases with increasing $k_{\rm M}\tau_{b}$ as show in the figure, before vanishing entirely when $k_{\rm M}\tau_{b}>{\mathcal O}(1)$.

\subsubsection{Small-wave number density fluctuations: }
\label{s:Sq}

\begin{figure}
\centerline{\includegraphics[width=8cm]{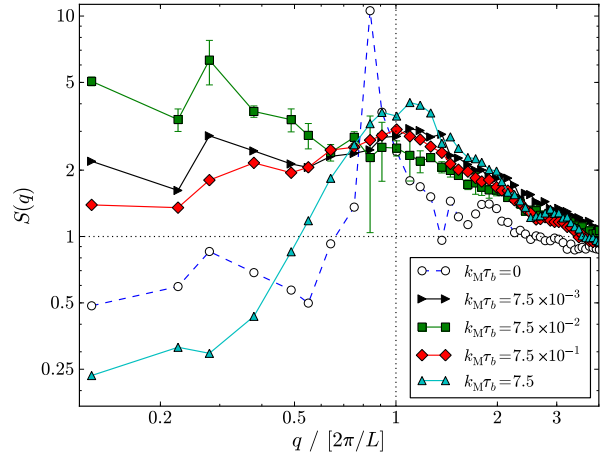}}
\caption{(Color online) Static structure factor $S(q)$ for $k_{\rm A}/k_{\rm D}=40$ and the motor speeds given in the key, for system sizes $8L\times8L$. For clarity only error bars for $k_{\rm M}\tau_{b}=0.075$ are shown; others are comparable. The horizontal and vertical dotted lines corresponds to $S(q)=1$ and $q=2\pi/L$ respectively.}
\label{f:Sq}
\end{figure}

Although useful for considering connectivity, the cluster distribution $P(n_{\rm c})$ gives no information regarding spatial distribution of clusters, and in particular exhibits no signature corresponding to the large void formation evident in snapshot of Fig.~\ref{f:snapshots}(d) for high $k_{\rm A}/k_{\rm D}$ and $k_{\rm D}\tau_{b}\stackrel{<}{\scriptstyle\sim}k_{\rm M}\tau_{b}\stackrel{<}{\scriptstyle\sim}1$. To consider the distribution of mass centers it is useful to look at the static structure factor~$S(q)$ as defined in Sec.~\ref{s:passive}. Fig.~\ref{f:Sq} shows the variation of $S(q)$ with $k_{\rm M}\tau_{b}$ for fixed $k_{\rm A}/k_{\rm D}=40$. Focussing on the $S(q\rightarrow0)$ behavior reveals a non-monotonic dependence on motor speed: for fast motors $k_{\rm M}\tau_{b}>{\mathcal O}(1)$, or slow, non-processive motors $k_{\rm M}<k_{\rm D}$, $S(q)$ decays to some small value as $q\rightarrow0$. For intermediate $k_{\rm M}$, $S(q)$ increases, possibly divergently, with decreasing~$q$, which coincides with the void formation evident in snapshots.

The limited number of data points and poor statistics for small $q$ makes it difficult to characterize the behavior of $S(q)$ in this limit. Nonetheless some semi-quantitative observations can be made by fitting each $S(q)$ to the form $A+Bq^{c}+C{\rm e}^{-(q/[2\pi/L]-1)^{2}/D^{2}}$ over the range $q\leq2q_{\rm L}$, which provides a stable fit for all parameters considered. Although the fitted exponent $c$ is somewhat susceptible to finite size effects, we consistently observe values close to $c=2$ for low motor densities $k_{\rm A}/k_{\rm D}\approx1$, or very slow or fast motors, $k_{\rm M}\ll k_{\rm D}$ and $k_{\rm M}\tau_{b}\gg1$ respectively. For high $k_{\rm A}/k_{\rm D}$ and intermediate $k_{\rm M}\tau_{b}$, $c$ decreases and becomes negative. Although the {\em magnitude} of the fitted exponent depends strongly on system size, the {\em sign} does not, suggesting some form of small-wavenumber structure will persist for large system size.

A divergent $S(q)$ was predicted from hydrodynamic equations of active nematic systems with an exponent $c=-2$\cite{Ramaswamy2003}, and was interpreted as emergent directed motion deriving from the coupling of elastic modes with the breaking of time-reversal invariance in driven states\cite{Ramaswamy2010}. We shall later argue for a form of emergent directed motion in our system when discussing the dynamics in Sec.~\ref{s:dynamics}, which derives however from the polar coupling between motors and filaments. It is possible that emergent directed motion drives the small wavenumber density fluctuations in both systems. As described in Sec.~\ref{s:superdiff}, the parameter values for which exponents $c<0$ arise coincide with those for which superdiffusive mass transport is most pronounced. It is interesting to note that a causal link between superdiffusion and small wavenumber fluctuations has been hypothesized from systems lacking orientational degrees of freedom\cite{Head2010b}, in broad agreement with our findings.

%
%

\subsubsection{Nematic and polar ordering: }
\label{s:nematic}

\begin{figure}
\centerline{\includegraphics[width=8cm]{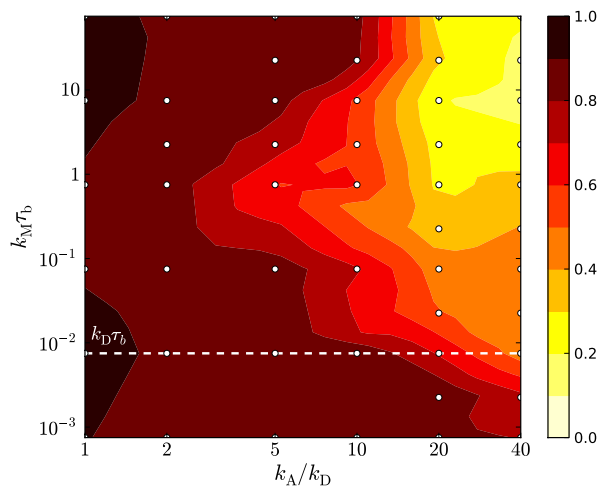}}
\caption{(Color online) Nematic ordering parameter $S_{2D}$ for $k_{\rm A}/k_{\rm D}$ and $k_{\rm M}\tau_{b}$ for fixed system size $4L\times4L$. In this figure, and in Fig.~\ref{f:OP_velpol} below, the white discs denote actual data points; linear interpolation within the enclosing triangle is used between points. The thick horizontal dashed white line corresponds to $k_{\rm M}=k_{\rm D}$.}
\label{f:OP_S2D}
\end{figure}

We now move beyond density fluctuations and consider the orientational degrees of freedom. Let $\hat{\bf p}^{\alpha}$ denote the unit polarity vector from the $[-]$-end to the $[+]$-end for each of the N filaments~$\alpha=1\ldots N$. From this can be defined the traceless nematic order tensor $Q_{ij}^{\alpha}=\hat{p}_{i}^{\alpha}\hat{p}_{j}^{\alpha}-\frac{1}{2}\delta_{ij}$. The two-dimensional nematic order parameter $S_{2D}$ is then defined by \mbox{$(S_{2D})^{2}=2\left\|\langle Q_{ij}\rangle\right\|^{2}$}, where $\langle Q_{ij}\rangle=N^{-1}\sum_{\alpha}Q_{ij}^{\alpha}$ and $\left\|\ldots\right\|^{2}$ denotes the Euclidean norm $\left\|A_{ij}\right\|=\sum_{i,j}A_{ij}^{2}$. $S_{2D}$ varies from 0 to~1, with 1 for perfect nematic ordering and 0 when there is no net preferred orientation, such as for an isotropic state, asters {\em etc}. A contour plot of $S_{2D}$ for $k_{\rm A}/k_{\rm D}$ and $k_{\rm M}\tau_{b}$ is given in Fig.~\ref{f:OP_S2D}, and confirms what is visible from snapshots in Fig.~\ref{f:snapshots}, namely that nematic ordering breaks down for $k_{\rm A}/k_{\rm D}\stackrel{>}{\scriptstyle\sim}15$ and processive motors with $k_{\rm M}\gg k_{\rm D}$. This data is for the smallest system size, for which we have most data points; the effects of system size have been checked and confirms the crossover to low $S_{2D}$ is robust. The value of $S_{2D}$ may be approaching zero for points in the upper right of this plot, but the poor statistics and limited number of points rules out extrapolation to the infinite system-size limit.

The nematic order parameter $S_{2D}$ tell us nothing about correlations in the {\em polarity} of filaments. To consider correlations in filament polarity, we must go beyond this nematic description and consider order parameters that respect reverses in polarity \mbox{$\hat{\bf p}\rightarrow-\hat{\bf p}$}. Firstly we note that the mean polarity $\langle\hat{\bf p}\rangle=N^{-1}\sum_{\alpha}\hat{\bf p}^{\alpha}$ vanishes for the entire parameter space considered here. This is in agreement with the theoretical prediction that states with non-zero mean filament polarity are only stable if the motors are polar, {\em i.e.} have attachment rates that depends on the relative filament polarity\cite{Ahmadi2005}, unlike the apolar motors considered here for which the attachment rate depends only on separation.

It is natural to consider spatial correlations in polarity along directions parallel to the filament axis separately to those in perpendicular directions. We therefore define the perpendicular polarity correlation function $C^{\perp}_{pp}(r)$ as

\begin{eqnarray}
C_{pp}^{\perp}(r)
&=&
\frac{
\sum_{\alpha,\beta}\hat{\bf p}^{\alpha}\cdot\hat{\bf p}^{\beta}\delta(r-|{\bf x}^{\alpha}-{\bf x}^{\beta}|)\sin^{2}\theta
}{
\sum_{\alpha,\beta}\delta(r-|{\bf x}^{\alpha}-{\bf x}^{\beta}|)\sin^{2}\theta
}
\nonumber\\
\label{e:cpp}
\end{eqnarray}

\begin{figure}[bp]
\centerline{\includegraphics[width=8cm]{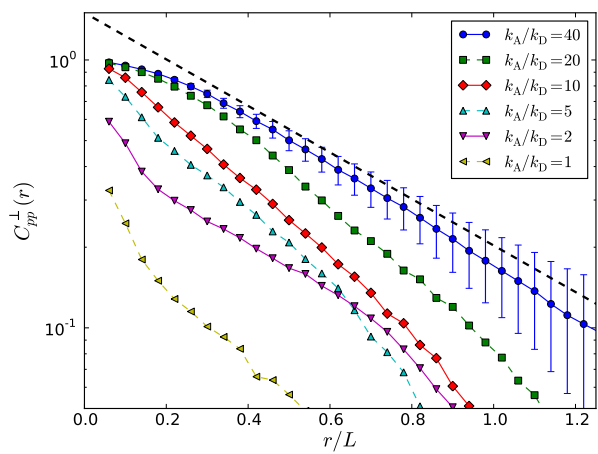}}
\caption{(Color online) The transverse polarity correlation function $C^{\perp}_{pp}(r)$ for $k_{\rm M}\tau_{b}=0.075$, system size $4L\times4L$ and the attachment rates given in the key. For clarity only errors bars for $k_{\rm A}/k_{\rm D}=40$ are shown; the others are comparable or smaller. The thick dashed line is proportional to $e^{-r/[L/2]}$.}
\label{f:polarityTrans}
\end{figure}

\noindent{}where $\theta$ is the angle between $\hat{\bf p}^{\alpha}$ and ${\bf x}^{\beta}-{\bf x}^{\alpha}$ ($\hat{\bf p}^{\beta}$ could equally have been chosen due to the symmetry of Eq.~(\ref{e:cpp})). The corresponding parallel function $C_{pp}^{\parallel}(r)$ is defined analogously, with the $\sin^{2}\theta$ weighting factors replaced by $\cos^{2}\theta$. Both projections exhibit approximate exponential decay with~$r$, with longer-range correlations for higher motor densities $k_{\rm A}/k_{\rm D}$, as demonstrated in Fig.~\ref{f:polarityTrans}. These correlations become shorter range as $k_{\rm M}\rightarrow0$, confirming the smooth approach to the passive limit $k_{\rm M}=0$ when all polarity correlations trivially vanish. It should be noted that significant polarity correlations are not mutually exclusive with nematic ordering. With regards to finite-size effects, the data shows no apparent trend with system size for all of the parameter space except for the high $k_{\rm A}/k_{\rm D}$ and $k_{\rm D}\tau_{b}\stackrel{<}{\scriptstyle\sim}k_{\rm M}\tau_{b}\stackrel{<}{\scriptstyle\sim}1$ when extended polar clusters arise, for which the range of polarity correlations was still increasing for the largest system sizes simulated. Finally, we never observe significant correlations on lengths larger than the filament length~$L$, which will be discussed in Sec.~\ref{s:discussion} in relation to hydrodynamic modelling.


%
%
\subsection{Dynamics}
\label{s:dynamics}

\begin{figure}
\centerline{\includegraphics[width=8cm]{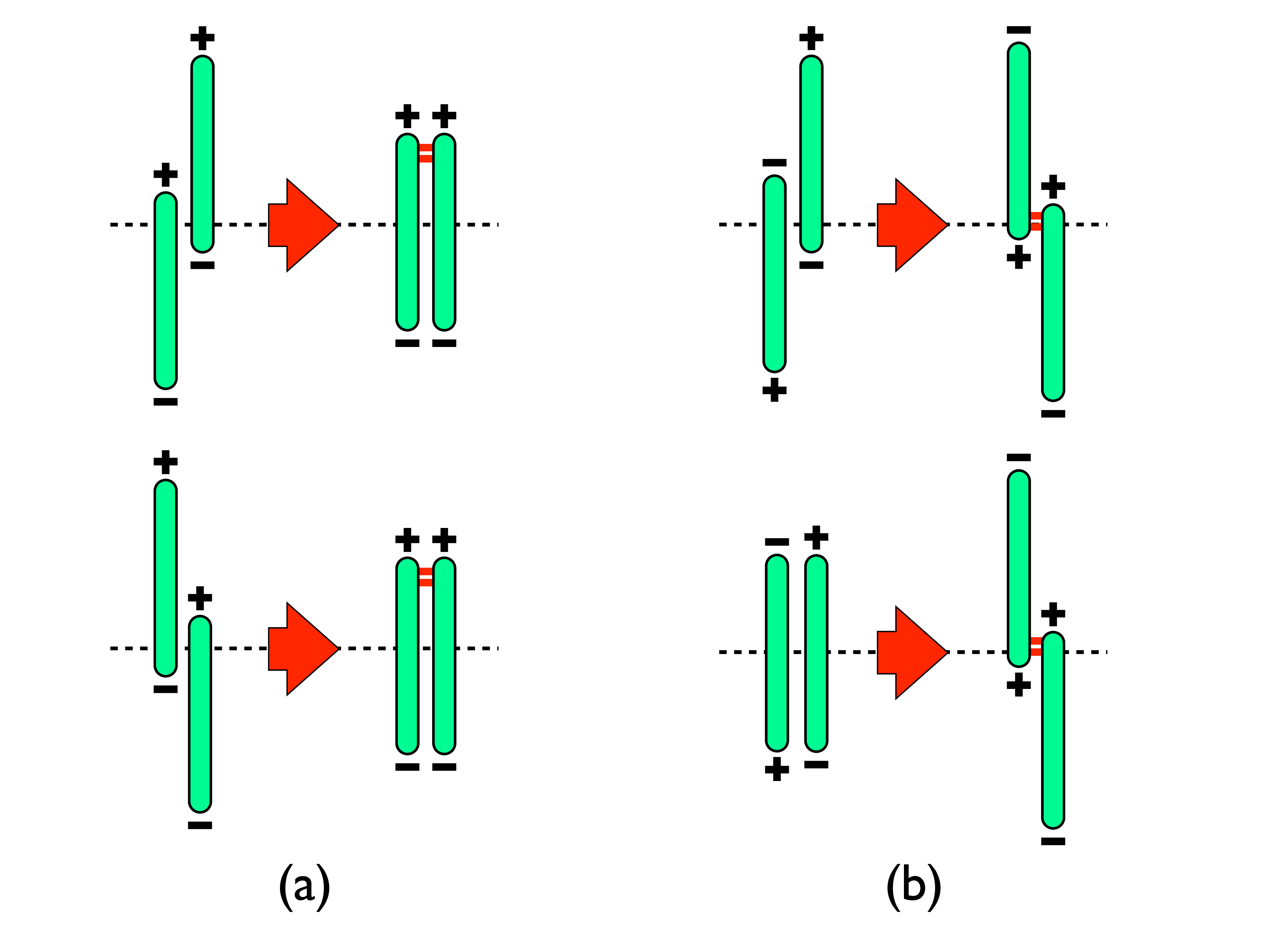}}
\caption{(Color online) Emergence of directional motion on the one-filament level under the action of motors, denoted by the short horizontal lines. (a)~Net zero displacement for parallel filaments. (b)~Net displacement towards the $[-]$--end for both filaments in an anti--parallel pairing.}
\label{f:Emergence_VelPolCorrn_noGradient}
\end{figure}

Motor motion represents an energy flux driving relative filament motion in violation of the principles of thermodynamic equilibrium, and might therefore be expected to result in anomalous  transport properties. Data confirming this expectation is provided in this section. The key observation underlying these findings is that the coupling between motors and filaments is polar, in that motor heads move strictly towards filament $[+]$-ends. Filaments are thus expected to move in the direction of their $[-]$-ends in response to the motor forces, as seen by considering connected parallel and anti-parallel filament pairs as in Fig.~\ref{f:Emergence_VelPolCorrn_noGradient}. This effect also emerges from one-dimensional continuum modeling\cite{Kruse2000,Kruse2001}. Thus filaments may move persistently in a direction that is coupled with their polarity. This scenario is reminiscent of self-propelled particles that are known to typically exhibit enhanced mass diffusion, such as an anomalous scaling with time of displacements transverse to particle motion\cite{Tu1998,Toner1998,Chate2008} or a longitudinal diffusion constant increased by the activity\cite{Baskaran2008}. In this light, anomalous transport should also be expected here.

\begin{figure}
\centerline{\includegraphics[width=8cm]{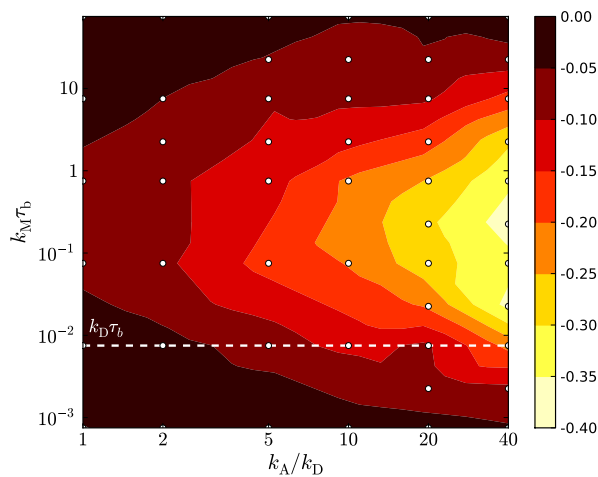}}
\caption{(Color online) Correlation between individual filament velocity and its polarity, $\langle{\bf v}\cdot\hat{\bf p}\rangle/v$, versus $k_{\rm A}/k_{\rm D}$ and $k_{\rm M}\tau_{b}$. As this quantity is relatively insensitive to system size, the largest available system was selected for each point.}
\label{f:OP_velpol}
\end{figure}

To confirm the correlations between filament motion and polarity, the filament velocity must be defined, for which it is necessary to average over a finite time interval since there is no instantaneous velocity in Brownian dynamics. We define the velocity of filament $\alpha$ at time $t$ to be ${\bf v}^{\alpha}=[{\bf x}^{\alpha}(t+\Delta t)-{\bf x}^{\alpha}(t)]/\Delta t$ where $\Delta t\approx65\tau_{b}$, corresponding to trajectories spanning a few monomer diameters. Correlations between polarity and velocity are then easily quantified as $\langle{\bf v}\cdot\hat{\bf p}\rangle/v$ with \mbox{$v=\sqrt{\langle{\bf v}\cdot{\bf v}\rangle}$} the mean filament speed. This is plotted in Fig.~\ref{f:OP_velpol}, and shows higher correlations for large $k_{\rm A}/k_{\rm D}$ and $k_{\rm M}\tau_{b}\sim10^{-2}-10^{-1}$, which we will show below is also roughly where the degree of anomalous mass transport is highest. Note also that the correlations are negative, confirming filaments move in the direction of their $[-]$-ends. 

\subsubsection{Super-diffusive mass transport: }
\label{s:superdiff}

\begin{figure}
\centerline{\includegraphics[width=8cm]{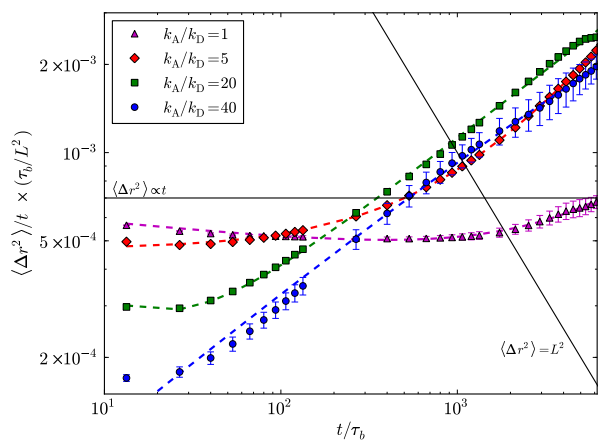}}
\caption{(Color online) Mean squared displacements divided by time, $\langle\Delta r^{2}\rangle/t$, so that normal diffusive scaling $\langle\Delta r^{2}\rangle\propto t$ shows up as a horizontal line. Here $k_{\rm M}\tau_{b}=0.075$, the system size was $8L\times8L$ and the $k_{\rm A}/k_{\rm D}$ are given in the legend. The thick dashed lines are fits to $C_{1}t^{c_{1}}+C_{2}t^{c_{2}}$ for each set of data points. The black horizontal line is to guide the eye, whereas the diagonal black line corresponds (on these axes) to $\langle\Delta r^{2}\rangle=L^{2}$ and confirms trajectories exceed the filament length for these data sets.}
\label{f:msd_kM1em2}
\end{figure}

Mass transport is conveniently quantified by the mean squared displacements of filament centers, $\langle\Delta r^{2}(t)\rangle$, where for simplicity we average over filament polarity. For the available time window, which typically corresponds to displacements from a fraction of $b$ to a few~$L$, a slow, diffusive or sub-diffusive region is observed at small times, sometimes followed by a crossover to super-diffusive scaling at later times in which $\langle\Delta r^{2}(t)\rangle\sim t^{\alpha}$ with $\alpha>1$. See Fig.~\ref{f:msd_kM1em2} for some examples. The variation of these curves with system size is weak or non-existent. As with the passive case in Sec.~\ref{s:passive}, we presume an eventual crossover to normal diffusion with $\alpha=1$ for lag times exceeding our achievable time window.

\begin{figure}
\centerline{\includegraphics[width=8cm]{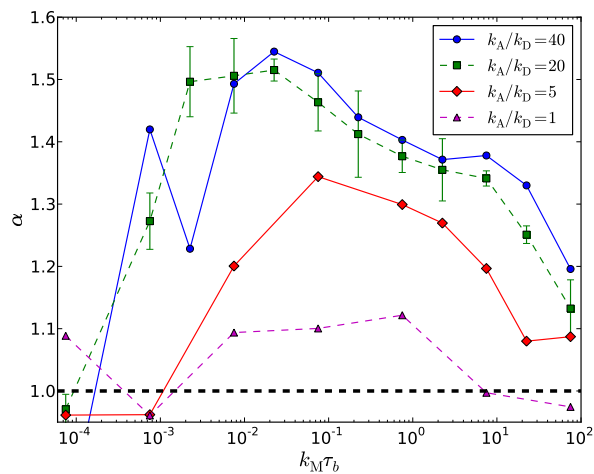}}
\caption{(Color online) Scaling of the mean squared displacement with time~$t$, {\em i.e.}~$\alpha=\partial\ln\langle\Delta r^{2}(t)\rangle/\partial\ln t$, at the point when $\langle\Delta r^{2}\rangle=L^{2}$, versus $k_{\rm M}\tau_{b}$ for the $k_{\rm A}/k_{\rm D}$ given in the key. For clarity only error bars for $k_{\rm A}/k_{\rm D}=20$ are shown; others are comparable}
\label{f:msdExp}
\end{figure}

To quantify the extent of deviation from normal diffusion as a function of the microscopic parameters, it is convenient to condense the variation of logarithmic slope $\alpha$ over the data window into a single scalar. To do this, we first smooth the data by fitting each curve to the sum of two power laws, $\langle\Delta r^{2}\rangle=C_{1}t^{c_{1}}+C_{2}t^{c_{2}}$, which gives a reasonable fit in all cases. We then take the logarithmic slope of this fit at the point when $\langle\Delta r^{2}\rangle=L^{2}$. The result is given in Fig.~\ref{f:msdExp}, and shows a decrease in anomalous diffusion away from the peak value at high $k_{\rm A}/k_{\rm D}$ and motor speeds roughly in the range $k_{\rm D}\tau_{b}\stackrel{<}{\scriptstyle\sim}k_{\rm M}\tau_{b}\stackrel{<}{\scriptstyle\sim}1$. Choosing a different point along the curve to extract the slope results in different values but similar trends.

A striking feature of Fig.~\ref{f:msdExp} is the non-monotonicity of the degree of anomalous diffusion with motor speed~$k_{\rm M}$, which approaches normal diffusion for $k_{\rm M}\tau_{b}\gg1$. The reason is not hard to find. As evident in Fig.~\ref{f:snapshots}(b), when motor motion dominates over thermal diffusion, they rapidly reach the filament's $[+]$-end where they then stall. Indeed, the fraction of motors with at least one head attached to a $[+]$-end never drops below $80\%$ for $k_{\rm M}\tau_{b}\geq1$. Since such heads can no longer move, they do not drive filament separation and the net motor activity decreases, restoring normal diffusion.

\subsubsection{Velocity correlations: }

\begin{figure}
\centerline{\includegraphics[width=8cm]{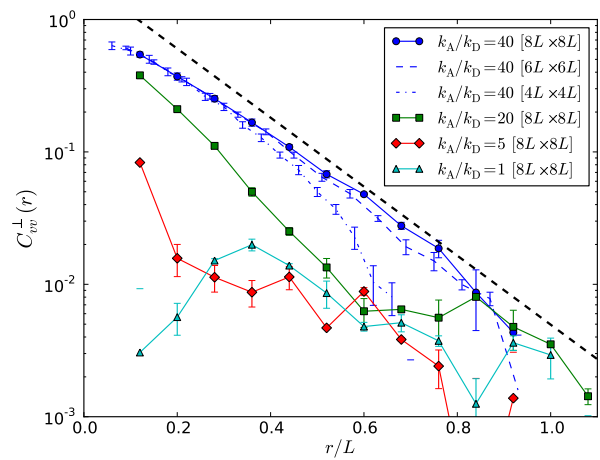}}
\caption{(Color online) Velocity correlations $C_{vv}^{\perp}(r)$ perpendicular to the filament axis, for $k_{\rm M}\tau_{b}=0.075$ and the $k_{\rm A}/k_{\rm D}$ and system sizes given in the legend. The thick dashed line is~$\propto e^{-r/[L/6]}$.}
\label{f:velCorrn}
\end{figure}

Using the same definition of filament velocity ${\bf v}^{\alpha}$ as above, it is possible to calculate spatial correlations in velocities projected parallel $C_{vv}^{\parallel}(r)$ and perpendicular $C_{vv}^{\perp}(r)$ to the filament axis. Here, $C_{vv}^{\parallel,\perp}(r)$ are defined analogously to the polarity correlations in Eq.~(\ref{e:cpp}), except with the $\hat{\bf p}^{\alpha}$ replaced by ${\bf v}^{\alpha}/v$ with $v$ the mean filament speed as before. Examples are given in Fig.~\ref{f:velCorrn}, demonstrating a growing range of correlated motion as the motor density increases. This trend was observed throughout the parameter space considered.

Also shown Fig.~\ref{f:velCorrn} is an example of the variation with system size for the highest $k_{\rm A}/k_{\rm D}$, which demonstrates convergence for the two largest system sizes considered. The decay is exponential, approximately of the form ${\rm e}^{-r/[L/6]}$ and thus becomes negligibly small on lengths on the order of the filament length~$L$. We never observe correlations decaying on lengths larger than~$L$, and conclude velocities on lengths much larger than the filament length are uncorrelated, a point that is discussed in Sec.~\ref{s:discussion}.


%
%
\section{Discussion}
\label{s:discussion}

\begin{figure}
\centerline{\includegraphics[width=8cm]{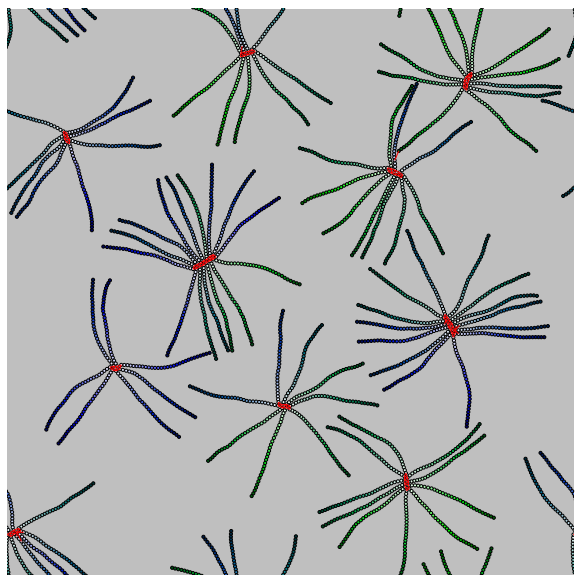}}
\caption{(Color online) Example of asters at low density with area fraction $\phi\approx5.3\%$. The system parameters were $k_{\rm A}/k_{\rm D}=250$, $k_{\rm M}\tau_{b}=75$. Motors are short straight lines concentrated at filament $[+]$-ends.}
\label{f:LowDenAster}
\end{figure}

It is apparent from the results described above that for the considered density of $\phi\approx21\%$, structures similar to the asters and vortices in the microtubule experiments\cite{Nedelec1997,Surrey2001} are never reproduced. The absence of asters may be a simple matter of density. The end-bound clusters formed by rapid motors in Fig.~\ref{f:snapshots}(b) are not able to form circular structures due to steric hinderance with nearby clusters. Lowering the filament density removes this effect, permitting full asters consisting of a single layer of filaments to form, as demonstrated in Fig.~\ref{f:LowDenAster}. However, vortices did not arise at lower densities, even when increased motor detachment rates at $[+]$-ends was included. We note that a continuum model~\cite{Sankararaman2004} that extended an earlier version~\cite{Lee2001} to include, amongst other features, a form of steric hinderance between filaments, favored asters over vortices relative to the earlier work, suggesting excluded volume may also be a factor. Here, we expand upon the relationship between our numerical findings and the related theory and experiments that has already been touched upon, with an emphasis on possible reasons for the non-observation of vortices.

%
%

\subsection{Comparison to experiments}

In their {\em in vitro} actin-myosin experiments, Backouche {\em et al.} claimed that active structure formation was only possible with the addition of a small concentration of the passive crosslinker fascin\cite{Backouche2006}. A small fraction of passive crosslinks between adjacent filaments has also been predicted to significantly increase their rate of alignment by mobile motors\cite{Ziebert2009}. Since no passive crosslinkers were included in our simulation, this might be a factor in the qualitatively different structures found here. Furthermore, fascin is a bundling protein that promotes the formation of {\em polar} actin bundles\cite{Vignjevic2006,Ishikawa2003}, thus its dominant role may in fact be to permit non-zero mean polarity. If so, it may play the role of the polar motor clusters that preferentially bind to parallel (as opposed to anti-parallel) filaments, which were shown theoretically to stabilize homogenous polarized states and produce a correspondingly richer stability diagram\cite{Ahmadi2005}. However, there is no reason to suppose that polarity-dependent binding of active or passive components played a role in the microtubule experiments, and was not included in the associated simulations\cite{Nedelec1997,Surrey2001}, suggesting this is not a reason for the non-emergence of vortices in our work.

Apart from molecular motors, a second form of non-equilibrium activity in biofilament gels is the spontaneous translation of filament center-of-mass {\em via} unequal monomer addition rates at opposite ends, known as treadmilling and prevalent {\em in vivo}\cite{BrayBook}. If present, this would place the system into the class of self-propelled particles, for which a richer variety of structure and dynamics is expected\cite{Ramaswamy2010,Tu1998,Toner1998,Chate2008,Baskaran2008,Peruiani2007,Kraikivski2006}. However, treadmilling was thought not to play a role in the actin-myosin experiments\cite{Backouche2006}, and was inhibited to some extent by taxol in the microtubule experiments (although microtubule growth {\em was} present)\cite{Nedelec1997,Surrey2001}. Until the role of treadmilling or filament growth is categorically denied experimentally this remains a possible missing factor, but without further experimental guidance we can merely speculate on its possible role. Including such features numerically should be straightforward, and indeed has already been performed for simulations of branching networks~\cite{Carlsson2001,Carlsson2003,Burroughs2007}.

Finally, the role of dimensionality should not be overlooked. Our simulations are strictly two-dimensional, whereas the actin-mysoin experiments were three-dimensional\cite{Backouche2006} and the microtubule experiments were performed in a thin $5\mu m$ chamber that enforces almost-parallel filament orientation while facilitating overlap, which can be regarded as a quasi-2D system\cite{Nedelec1997,Surrey2001} (the associated simulations were two-dimensional but without excluded volume, thus also permitting free overlap). The interaction terms in the hydrodynamic and mesoscale models were based on a three-dimensional kernel\cite{deGennesProst,DoiEdwards}. Thus the excluded volume interactions included in our simulations may be far more strict than the experiments or models, possibly having an inhibitory effect on structure formation. To numerically probe higher dimensionality is computationally expensive but should be possible with a restricted sampling of parameter space.

\subsection{Comparison to hydrodynamic models}

An assumption common to many of the analytical models is that the velocity and polarity fields are hydrodynamic, in the sense that they have long-wavelength components extending across lengths much larger than the filament length~$L$. In contrast, as described in Secs.~\ref{s:statics} and \ref{s:dynamics} above, we never observe polarity or velocity correlations that decay on lengths larger than~$L$.
It is possible that long-wavelength correlations emerge at much higher filament densities, but we were not able to check this so far due to computational limitations. Alternatively, our numerical scheme may be oversimplified, in that momentum is not conserved by the solvent-filament interactions. Since the standard argument for the emergence of a hydrodynamic velocity field requires momentum conservation\cite{ChaikinLubensky}, this may explain its absence, but correcting for this omission numerically is difficult and will require {\em ad hoc} modification of the driving terms such as in Ref.\cite{Fiege2009}, or the incorporation of full hydrodynamic interactions\cite{Gompper2008}.

A slowly varying polarity field is assumed in all theoretical models and coarse-graining schemes\cite{Kruse2004,Kruse2005,Voituriez2005,Voituriez2006,Marenduzzo2007,Basu2008,Cates2008,Liverpool2003,Ziebert2004,Liverpool2004,Ziebert2005,Liverpool2005,Ahmadi2005,Ruhle2008,Bassetti2000,Lee2001,Kim2003,Sankararaman2004}. The short-ranged decay of the polarity correlations in our simulations therefore makes the comparison of the results difficult. This may be partly a question of time scale. For the density considered here, the rotational diffusion time for filaments to significantly change orientation can be large. This is particularly true when the nematic order parameter $S_{2D}$ is close to unity, when the orientation autocorrelation function decays by as little as 5\% over a production run. This may inhibit the formation of large, polarity-correlated regions. Nonetheless even in this region  of parameter space, motor-driven {\em translational} separation of filaments according to their polarity does occur. Furthermore when $S_{2D}\ll1$, filament rotation is substantially enhanced. The lack of long-wavelength polarity correlations is therefore somewhat of a mystery.

\begin{figure}
\centerline{\includegraphics[width=8cm]{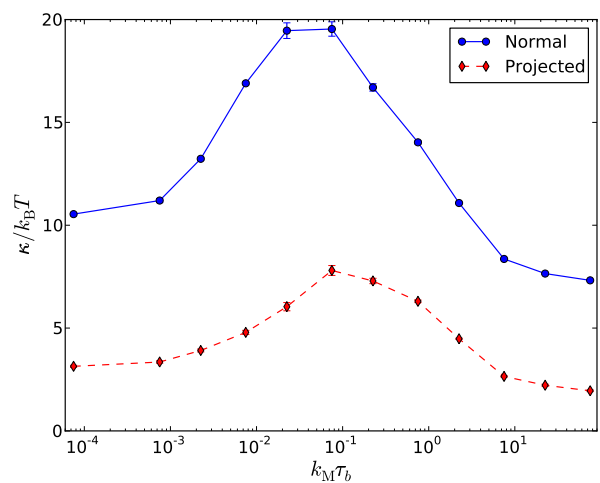}}
\caption{(Color online) Variation of the dipole moment with $k_{\rm M}$ for $k_{\rm A}/k_{\rm D}=40$ and system size $4L\times4L$ in steady state. The upper line gives the mean dipole moment $\kappa=\frac{1}{N_{(\alpha\beta)}}\sum_{(\alpha\beta)}{\bf F}^{\alpha\beta}\cdot({\bf x}^{\beta}-{\bf x}^{\alpha})$ as defined in the text, and the lower line gives the projected equivalent $\kappa^{\rm proj}=\frac{1}{2N_{(\alpha\beta)}}\sum_{(\alpha\beta)}\left\{({\bf F}^{\alpha\beta}\cdot\hat{\bf p}^{\alpha})(\Delta{\bf x}^{\alpha\beta}\cdot\hat{\bf p}^{\alpha})+({\bf F}^{\beta\alpha}\cdot\hat{\bf p}^{\beta})(\Delta{\bf x}^{\beta\alpha}\cdot\hat{\bf p}^{\beta})\right\}$ with $\hat{\bf p}^{\alpha}$, $\hat{\bf p}^{\beta}$ the polarity unit vectors for filaments $\alpha$ and $\beta$ resp.}
\label{f:dipole}
\end{figure}

The nematodynamic theories take as an input parameter the active stress generated on filaments, whose sign determines if this stress is contractile or extensile\cite{Kruse2004,Voituriez2006}. Our microscopic approach does not impose the sign of the force dipoles generated by motors, so instead it must be measured. The mean dipole moment acting between filament pairs is defined as $\kappa=\frac{1}{N_{(\alpha\beta)}}\sum_{(\alpha\beta)}{\bf F}^{\alpha\beta}\cdot({\bf x}^{\beta}-{\bf x}^{\alpha})$, where the sum is over all $N_{(\alpha\beta)}$ filaments pairs $(\alpha,\beta)$ connected by at least one motor, and ${\bf F}^{\alpha\beta}$ is the total force on $\alpha$ due to motors connecting $\alpha$ and~$\beta$. Thus $\kappa>0$ corresponds to contractile dipoles, and $\kappa<0$ to extensile ones. Preliminary results are given in Fig.~\ref{f:dipole}, demonstrating uniformly contractile stress $\kappa>0$ with a magnitude that reaches a maximum for  intermediate motor speeds $k_{\rm D}\tau_{b}<k_{\rm M}\tau_{b}<1$. An alternative measure in which both ${\bf F}^{\alpha\beta}$ and ${\bf x}^{\beta}-{\bf x}^{\alpha}$ are first projected along the filaments' axes before summing, as defined in the figure caption, follows the same trend. Thus our motor rules lead to contractile active stresses, theoretically predicted to give the richest behaviour\cite{Kruse2004,Voituriez2006}. There is also a slight but measurable reduction of $\approx0.4\%$ in the mean filament contour length when $\kappa$ is at its highest, confirming contractility.

%
%

\section*{Acknowledgements}
The authors would like to thank J. Padding, I G\"otze and J. Elgeti for useful discussions. Financial support of this project by the European Network of Excellence ``SoftComp'' though a joint postdoctoral fellowship for DAH is gratefully acknowledged.

%
%

\end{document}